\documentclass[twocolumn,english,aps,pra,showpacs]{revtex4}
\usepackage{amssymb}
\usepackage{amsmath}
\usepackage{colordvi}
\usepackage{amsthm}
\usepackage{epsfig}
\usepackage{graphicx}
\usepackage{subfigure}
\usepackage{color}

\def\avg(#1){\langle#1\rangle}

\def\be{\begin{equation}}
\def\ee{\end{equation}}
\def\bea{\begin{eqnarray}}
\def\eea{\end{eqnarray}}

\begin{document}

\title{\textbf{Dynamical Spin Squeezing via Higher Order Trotter-Suzuki Approximation}}
\author{Ji-Ying Zhang}
\author{Xiang-Fa Zhou}
\email{xfzhou@ustc.edu.cn}
\author{Guang-Can Guo}
\author{Zheng-Wei Zhou}
\email{zwzhou@ustc.edu.cn}
\affiliation{Key Laboratory of Quantum Information, University of Science and Technology of China, Chinese Academy of Sciences, Hefei, Anhui 230026, China}

\affiliation{Synergetic Innovation Center of Quantum Information and Quantum Physics, University of Science and Technology of China, Hefei,
Anhui 230026, China}

\begin{abstract}
 Here we provide a scheme of transforming the one-axis twisting Hamiltonian into a two-axis twisting Hamiltonian based on high-order Trotter-Suzuki approximation. Compared with the paper by Liu \textsl{et al}. [Phys. Rev. Lett. 107, 013601 (2011)], our method can reduce the number of controlling cycles from $1000$ to $50$. Moreover, it is also spin number independent and takes a shorter optimal evolution time as compared with the method of Shen \textsl{et al}.[Phys. Rev. A 87, 051801 (2013)]. The corresponding error analysis is also provided.
\end{abstract}

 \maketitle
 \section{Introduction}
Squeezed spin states \cite{Kitagawa,Heinzen,Wineland,Ma} are entangled quantum states of an ensemble of two-level (or spin-half) systems, and they play significant roles in quantum information science \cite{Zoller,Bigelow,Sorensen,Korbicz,Amico,Pezze,Horodecki,Guhne,Lee} and quantum metrology \cite{Heinzen,Wineland,Bollinger,Leibfried,Polzik,Cronin,Appel,Gross,Riedel}. People have made much progress in both theory and experiment over the past decades \cite{Ma,Gross,Riedel,Haine,Liu,Shen,Yin,Louchet,Julia}. Specifically, the recent experimental success of achieving the one-axis twisting (OAT) scheme in spinor Bose-Einstein condensates (BEC) using two chosen hyperfine states provides an ideal platform to implement such novel states in a highly controllable manner \cite{Gross,Riedel}.

As is well known, two-axis twisting (TAT) is capable of causing Heisenberg limited noise reduction to scale as $1/N$, better than the OAT, whose noise reduction limit scales as $1/N^{2/3}$ \cite{Kitagawa}. To realize better spin squeezing, several theoretical proposals have been presented to enhance the OAT spin squeezed states \cite{Liu,Shen,Chaudhry}. In one scheme \cite{Liu}, one applies a series of subtle Rabi pulses to the system with the purpose of transforming OAT into TAT. Due to a large number of pulses acting on the atoms, it's unavoidable to bring in accumulated noise and imperfection in control pulses. In another approach \cite{Shen}, only several pulses are needed to obtain much better squeezed spin states. However, to achieve the optimal squeezing it takes a long evolution time, which would be an obstacle in systems with short coherence time. Additionally, this scheme is also spin number dependent, so it naturally brings in certain difficulties when applied to some systems, such as ultracold atomic gases, in which we do not know the spin number $N$ exactly.

Here we propose a scheme following the idea of transforming OAT into TAT to enhance the performance of OAT. Compared with the method discussed in the paper of Ref. \cite{Liu}, pulse sequences based on Trotter-Suzuki (TS) expansion \cite{Suzuki} are proposed. To achieve this, we also introduce another kind of radio frequency (rf) pulses to realize the rotation around the $x$ axis apart from that around the $y$ axis \cite{Liu}. We note that the scheme can be generalized to implement pulse sequences based on any high order TS expansion within these experimentally available conditions. A numerical investigation of the scheme based on the $2$nd-order expansion indicates that only $50$ cycles are enough to obtain the ideal spin squeezed states, while more than $1000$ cycles are needed in \cite{Liu} to get the same results. So compared with the previous proposals \cite{Liu,Shen}, our idea can overcome their disadvantage to some extent. Moreover, we also provide the corresponding error analysis for a scheme using higher- order TS expansions.

\section{The schemes and pulse sequences}
To clarify our key point in this paper, we first briefly review the TS expansion theory \cite{Suzuki}. The standard $1$st- and $2$nd-order TS real decomposition of $e^{\alpha (P+Q)}$ (with the commutation relation $[P,Q]\neq 0$ in terms of operators $P$ and $Q$) for small $|\alpha| (|\alpha|\ll1)$ are
\begin{equation}
\begin{split}
e^{\alpha(P+Q)}&=e^{\alpha P}e^{\alpha Q}+O(\alpha^2)\\
e^{\alpha(P+Q)}&=S(\alpha)+O(\alpha^3)\\
&=e^{(\alpha/2)P}e^{\alpha Q}e^{(\alpha/2)P}+O(\alpha^3).
\end{split}
\label{(Eq.tro1&2)}
\end{equation}
For the $3$rd-order expansion, we begin with
\begin{equation}
e^{\alpha(P+Q)}=e^{s\alpha(P+Q)}e^{(1-2s)\alpha(P+Q)}e^{s\alpha(P+Q)}.
\label{(Eq.tro3.1)}
\end{equation}
The $3$rd-order symmetric approximation $S_3(\alpha)$ is given by
\begin{equation}
S_3(\alpha)=S(s\alpha)S((1-2s)\alpha)S(s\alpha)
\label{(Eq.tro3.2)}
\end{equation}
with the parameter $s=1/(2-2^{(1/3)})\simeq1.3512$. The $4$-th order expansion is the same as the $3$rd-order one, $S_4(\alpha)=S_3(\alpha)$ \cite{Suzuki}. In general, the $(2m-1)$th and $2m$th approximants, $S_{2m-1}(\alpha)$ and $S_{2m}(\alpha)$, are determined recursively as
\begin{equation}
\begin{split}
S_{2m-1}(\alpha)&=S_{2m}(\alpha)\\
&=S_{2m-3}(k_m\alpha)S_{2m-3}((1-2k_m)\alpha)S_{2m-3}(k_m\alpha)
\end{split}
\label{(Eq.tro.Higher)}
\end{equation}
with the parameter $k_m=(2-2^{1/(2m-1)})^{-1}$.

According to Refs \cite{Gross,Riedel}, the OAT Hamiltonian existing in two-component BEC controlled by coupling pulses can be written as
\begin{equation}
H=\chi J_z^2+ G(t)J_x+\Omega(t)J_y.
\label{(Eq.Hamiltonian)}
\end{equation}
Here $J_{\mu}=\sum_{k=1}^N\sigma_{\mu}^k/2$ is the collective angular momentum operator for the system with $N$ spins, $\mu=x,\,y,\,z$. $\chi$ indicates the nonlinear interaction strength between the atoms. $\Omega(t)$ and $G(t)$ are defined as the coupling pulse amplitudes. The model Hamiltonian in Eq.(5) is the so-called Lipkin-Meshkov-Glick model \cite{Lipkin}. Some aspects of this model have been discussed in Ref. \cite{Vidal}. Here and in the following, we assume $\Omega(t)=\Omega_{0}$ and $G(t)=G_{0}$ when the coupling pulses are switched on and $\Omega(t)=G(t)=0$ when they are turned off. Note that we will ignore the nonlinear interaction $\chi J_z^{2}$ during the time applying the coupling pulses, because the conditions $\chi N \ll \mid\Omega_{0}\mid $ and $\chi N \ll \mid G_{0}\mid$ are satisfied when the strong coupling Rabi pulses are switched on.

The terms $G(t)J_x$ and $\Omega(t)J_y$ can realize the rotation of $e^{-i\chi J_z^2t}$ around the $x$ and $y$ axis by angle $\theta\in(0,2\pi)$ with $\theta=\int_{-\infty}^\infty dt \Omega(t)(G(t))$. The rotation operator $R_{\theta}^{x(y)}$ is defined as
\begin{equation}
R_{\theta}^{x(y)}=e^{-i\theta J_x(J_y)}
\end{equation}
and it rotates $e^{-i\chi J_z^2t}$ as follows
\begin{equation}
\begin{split}
R_{-\theta}^{x(y)} e^{-i\chi J_z^2t}R_{\theta}^{x(y)}
=e^{-i\chi(J_z cos(\theta)+J_{y(x)}sin(\theta))^2t}.\\
\end{split}
\label{(Eq.Rotation_x&y)}
\end{equation}
Using this definition, the combination of $\theta=\pi/2$ and $\theta=-\pi/2$ is able to accomplish the following operations
\begin{equation}
\begin{split}
&R_{-\pi/2}^x e^{-i\chi J_z^2t} R_{\pi/2}^x=e^{-i\chi J_y^2t},\\
&R_{-\pi/2}^y e^{-i\chi J_z^2t} R_{\pi/2}^y=e^{-i\chi J_x^2t}.
\end{split}
\label{(Eq.pi/2Rotation)}
\end{equation}
From Eq. (\ref{(Eq.Rotation_x&y)}) we find out that the terms $e^{i\chi J_x^2t}$ and $e^{i\chi J_y^2t}$ can-not be realized directly with the Hamiltonian shown by Eq. (\ref{(Eq.Hamiltonian)}).

\begin{figure}
\includegraphics[width=1.0\columnwidth]{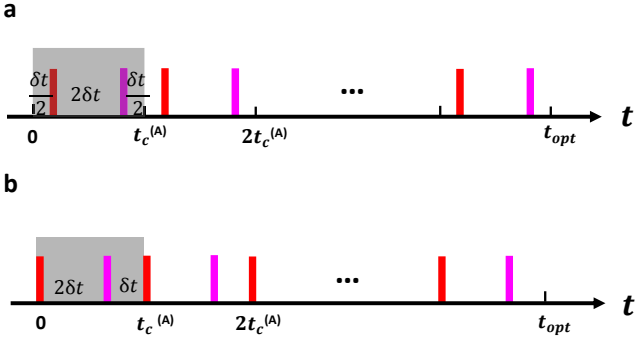}
  \caption{(color online). Schematic plot of the repeated pulses $\Omega_0$ vs time $t$ in arbitrary unit. (a) The whole pulse sequence of scheme A. One period from 0 to $t_c^{(A)}$ (shaded) consists of rotations $R_{\pi/2}^y$ (red pulse) and $R_{-\pi/2}^y$ (pink pulse). (b) The proposal in paper \cite{Liu}. Apart from the time at which applying the laser pulses, others are all the same with (a).}
\label{Fig.tro1&2.Pulse}
\end{figure}

To generate the TAT evolution $e^{-i\chi (J_x^2-J_y^2)t}$, we notice that $J^2$ is conserved during the dynamics. So up to a constant phase factor, we can write $e^{-i\tau (J_x^2-J_y^2)}$ as $e^{-i\tau (J^2+J_x^2-J_y^2)} = e^{-i\tau (2J_x^2+J_z^2)}$ with $\tau=\chi \delta t$ and $\delta t$ is a small time interval. Therefore the $1$st- and $2$nd-order expansion can be obtained as
\begin{equation}
\begin{split}
&e^{-i\tau (J_x^2-J_y^2)}\simeq e^{-i\tau J_z^2} e^{-i2\tau J_x^2}+ O((-i\tau )^2),\\
&e^{-i\tau (J_x^2-J_y^2)}\simeq e^{-i\frac{\tau}{2} J_z^2} e^{-i2\tau J_x^2}e^{-i\frac{\tau}{2} J_z^2} +O((-i\tau)^3).
\end{split}
\label{(Eq.tro1&2_expansion)}
\end{equation}
Equation (\ref{(Eq.tro1&2_expansion)}) tells us the rotation $R^y_{\pm \pi/2}$ is required to realize the evolution $e^{-i2\tau J_x^2}$, namely, after introducing the $\Omega(t)J_y$ pulse we can simulate the TAT based on Eq. (\ref{(Eq.Hamiltonian)}). The work based on TS $1$st-order expansion has been finished by Liu \textsl{et al}. \cite{Liu}, and the pulse sequence is shown in Fig. \ref{Fig.tro1&2.Pulse}(b).

Next we will provide the expansion scheme according to the TS $2$nd-order expansion theory shown in Eq. (\ref{(Eq.tro1&2_expansion)}). Figure \ref{Fig.tro1&2.Pulse}(a) shows the pulse sequences of our scheme A. Within each period, two strong pulses $R^y_{\pi/2}$ and $R^y_{-\pi/2}$ are employed at time $\delta t/2+Mt_c^{(A)}$ and $5\delta t/2+Mt_c^{(A)}$ respectively to realize the rotation shown in Eq. (\ref{(Eq.pi/2Rotation)}). Here $t^{(A)}_c=3\delta t$ is the time interval of a single period and $M=0,\,1,\,\cdots,\,N_c-1$ with $N_c$ the number of total period. In this case we have $N_p^{(A)}=2$, where $N_p^{(A)}$ is defined as the pulse number added in each period.

Without the controlling pulses, the dynamics of the system is determined by the Hamiltonian $H=\chi J^2_z$, so the evolution operator for one single period $U_1$ is
\begin{equation}
\begin{split}
U_1&=e^{-i \frac{\tau}{2} J_z^2} R_{-\pi/2}^y e^{-i 2\tau  J_z^2}R_{\pi/2}^y e^{-i \frac{\tau}{2} J_z^2}\\
&= e^{-i\tau(2J_x^2+J_z^2)}+O((-i\tau)^3).
\end{split}
\label{(Eq.U1.eff)}
\end{equation}

If we bring in $N_c$ periods the same as the one described above during a fixed interested time, the complete time evolution operator at time instant $t=N_ct_c^{(A)}$ is written as
\begin{equation}
\begin{split}
U_1^{N_c} &\simeq e^{-iN_c\tau(2J_x^2+J_z^2)}=e^{-i \frac{2J_x^2+J_z^2}{3}\chi t}\\
&\simeq e^{-i \frac{J_x^2-J_y^2}{3}\chi t}.
\end{split}
\label{(Eq.U1n.eff)}
\end{equation}
From Eq. (\ref{(Eq.U1n.eff)}) we find that the effective Hamiltonian of the system is $H_{eff}^{(A)}=\frac{\chi}{3}(J_x^2-J_y^2)$. Hence to realize the TAT evolution $U_{TAT}=e^{-i\chi(J_x^2-J_y^2)t_{opt}}$, with $t_{opt}$ the time when the optimal squeezing state is achieved, our scheme A takes the total time $3t_{opt}$.

To proceed with our scheme B, let us refer to the TS $3$rd-order expansion formula Eq. (\ref{(Eq.tro3.2)}). Unfortunately, there exists a term $S((1-2s)\alpha)$ on its right side, which can not be realized directly since $2s-1>0$. To solve this, we go back to Eq. (\ref{(Eq.tro3.1)}) and transform $e^{(1-2s)\alpha (P+Q)}$ to $e^{(2s-1)\alpha (-P-Q)}$. Taking into account the property of $J^2$, we obtain
\begin{equation}
\begin{split}
&e^{-i\tau (J_x^2-J_y^2)}\\
\simeq&e^{-is\tau(2J_x^2+J_z^2)}e^{-i(2s-1)\tau (2J_y^2+J_z^2)}
e^{-is\tau(2J_x^2+J_z^2)}.
\end{split}
\end{equation}
Therefore, following the same routine, we have the final result as
\begin{equation}
\begin{split}
&e^{-i\tau(J_x^2-J_y^2)}\\
\simeq &e^{-i\frac{s}{2}\tau J_z^2} e^{-i2s\tau J_x^2}e^{-i\frac{(3s-1)}{2}\tau J_z^2}
e^{-i2(2s-1)\tau J_y^2}\\\times &e^{-i\frac{(3s-1)}{2}\tau J_z^2}e^{-i2s\tau J_x^2}e^{-i\frac{s}{2}\tau J_z^2}+O((-i\tau)^4).
\end{split}
\label{(Eq.tro3_expansion)}
\end{equation}

\begin{figure}
  \includegraphics[width=1.0 \columnwidth]{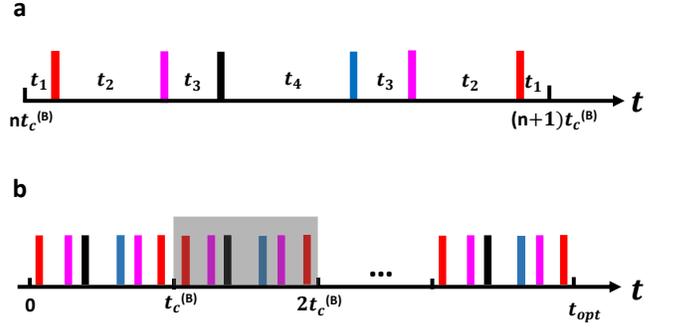} 
  \caption{(color online) An illustration of the pulse sequences for our scheme B. The red, pink, black and blue rectangles correspond to the $\pi/2$, $-\pi/2$ pulse around the $y$ axis, and $\pi/2$, $-\pi/2$ pulse around the $x$ axis respectively. (a) The pulse sequence for one single period, including a total of six pulses. (b) A series of pulse sequence periods. One period, from $t_c^{(B)}$ to $2t_c^{(B)}$, is shaded.}
\label{Fig.tro3.Pulse}
\end{figure}

From Eq. (\ref{(Eq.tro3_expansion)}), we find that both the coupling pulses $G(t)J_x$ and $\Omega(t)J_y$ are needed to implement the evolution $e^{-i2s\tau J_x^2}$ and $e^{-i2(2s-1)\tau J_y^2}$. Figure \ref{Fig.tro3.Pulse}(a) shows the corresponding pulse sequences within one single period: two pulses $R_{\pi/2}^y$ and $R_{-\pi/2}^y$ are employed at time $T_1+M t_c^{(B)}$ and $T_2+M t_c^{(B)}$, respectively, to implement a rotation around the $y$ axis; then another two pulses $R_{\pi/2}^x$ and $R_{-\pi/2}^x$ are added at time $T_3+M t_c^{(B)}$ and $T_4+M t_c^{(B)}$; and finally, a $y$ rotation is applied again with the pulses added at time $T_5+M t_c^{(B)}$ and $T_6+M t_c^{(B)}$, respectively. Here $T_{\nu}=\sum_{i=1}^{\nu}t_i$, $t_i=t_{8-i}$, with $t_1=s\delta t/2$, $t_2=2s\delta t$, $t_3=(3s-1)\delta t/2$, and $t_4=2(2s-1)\delta t$. We note that the duration time of one single period is $t_c^{(B)}=(12s-3)\delta t\simeq13.2\delta t$. In scheme B, the pulse number needed in one single period is $N_p^{(B)}=6$.

Following the similar way of getting the result of our scheme A, we conclude that the effective evolution of our scheme B is
\begin{equation}
U_2^{N_c}\simeq e^{-i\frac{J_x^2-J_y^2}{12s-3}\chi t}
\end{equation}
with an effective Hamiltonian $H_{eff}^{(B)}=\frac{\chi}{12s-3}(J_x^2-J_y^2)$. Furthermore, in this case the evolution time arriving at the optimal squeezing is $(12s-3)t_{opt}$. We note that the above method can be generalized to implement the TAT Hamiltonian based on any higher-order TS expansion.

\begin{figure}
 \includegraphics[width=1.0 \columnwidth]{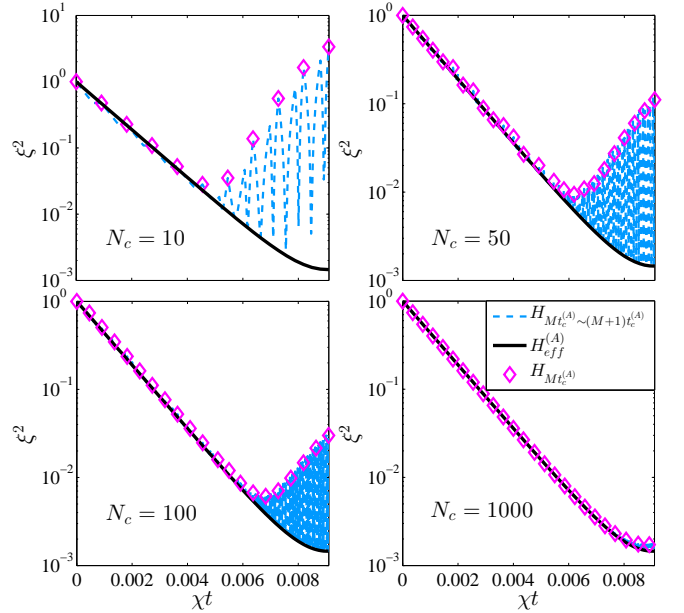} 
  \caption{(Color online). The resulting spin squeezing vs the evolution time based on the scheme in Ref. \cite{Liu} for the system with 1250 atoms. The magenta diamonds, the dashed light blue lines, and the solid black lines display the result for the case where $Mt_c^{(A)}$ is considered, every time instant from $Mt_c^{(A)}$ to $(M+1)t_c^{(A)}$ $(M=0,\,1,\,2,\,\cdots,\,N_c-1)$ is taken into account, and where $H_{eff}^{(A)}$ is concerned. $N_c$ is the number of total periods.}
\label{Fig.tro1.ntc&non_ntc}
\end{figure}

\begin{figure}
 \includegraphics[width=1.0 \columnwidth]{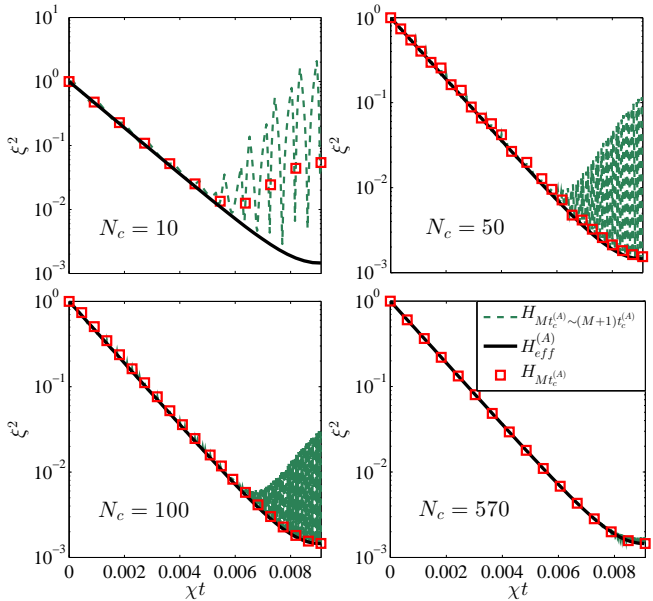} 
  \caption{(Color online). Same as in Fig. \ref{Fig.tro1.ntc&non_ntc}, except this time scheme A is considered. The red squares are used for the result at time $Mt_c^{(A)}$. The dashed dark green lines show the results when every time instant from $Mt_c^{(A)}$ to $(M+1)t_c^{(A)}$ $(M=0,\,1,\,2,\,\cdots,\,N_c-1)$ is considered.}
\label{Fig.tro2.ntc&non_ntc}
\end{figure}

\begin{figure}
 \includegraphics[width=1.0 \columnwidth]{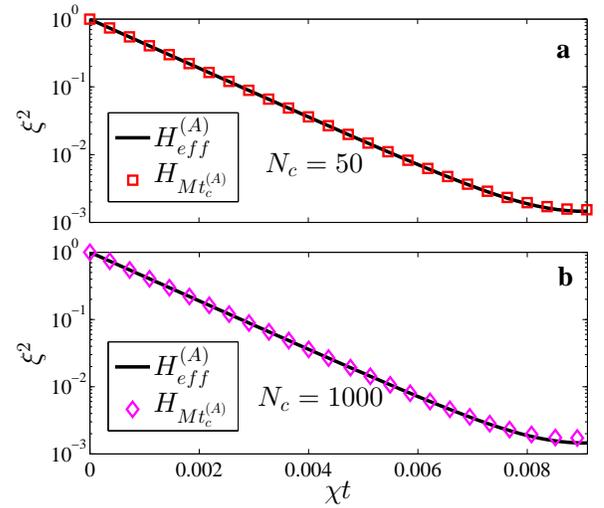} 
  \caption{(Color online) The numerical result of spin-squeezing parameter of (a) our scheme A and (b) the proposal in \cite{Liu}. The red squares and the magenta diamonds represent the results at time $Mt_c^{(A)}$ using the dynamics controlling pulses in the schemes. The black lines denote the results derived from the ideal effective TAT Hamiltonian $H_{eff}^{(A)}$ directly. Here $N=1250$ is used.}
\label{Fig.tro1&2.ntc}
\end{figure}

\begin{figure}
 \includegraphics[width=1.0 \columnwidth]{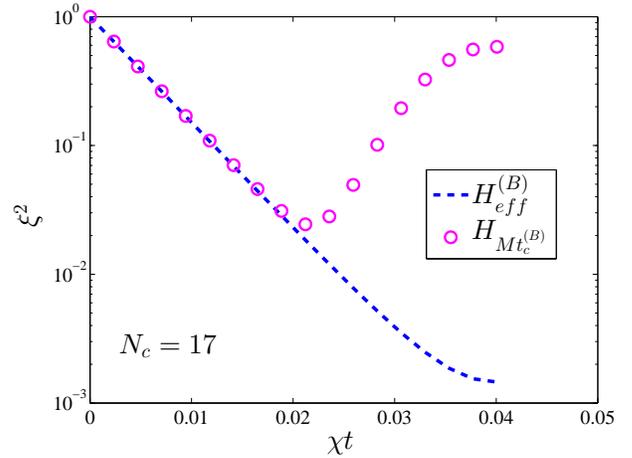}  
  \caption{(Color online) Squeezing parameter $\xi^2$ as a function of evolution time $\chi t$ calculated with $N=1250$ and $N_c=17$ ($N_p\approx100$). The magenta circles represent the values derived from the actual process of the pulse sequence shown in Fig. \ref{Fig.tro3.Pulse}(b) and the dashed blue line indicates the result for $H_{eff}^{(B)}$.}
\label{Fig.tro3.ntc}
\end{figure}

\section{The result and analysis}
To get the numerical result, we follow Kitagawa and Ueda's criteria that choose the squeezing parameter $\xi^2=\frac{2(\triangle J_\bot)_{min}^2}{J}$ as the measurement of the squeezing, where $(\triangle J_\bot)_{min}^2$ is the smallest variance normal to the mean spin vector and $J=N/2$ is the total spin of the system. The initial state we choose is $|J,J\rangle$, where all the spins are polarized along the $z$ axis. This state is favorable when the twisting Hamiltonian $H\propto J_x^2-J_y^2$.

It is shown in \cite{Liu} that for the scheme based on the $1$st-order expansion, $1000$ pulse pairs are enough to get the optimal spin squeezing with required precision. This result is obtained by taking into account every time instant in the time period from $Mt_c^{(A)}$ to $(M+1)t_c^{(A)}$ $(M=0,\,1,\,2,\,\cdots,\,N_c-1)$. That is, at every time instant $t$ satisfying $Mt_c^{(A)} \leq t \leq (M+1)t_c^{(A)}$, the approximated time evolution almost overlaps with the dynamics driven by the ideal TAT Hamiltonian. However, according to the theoretical analysis presented in Eqs. (\ref{(Eq.U1.eff)}) and (\ref{(Eq.U1n.eff)}), we notice that only the result at time instant $Mt_c^{(A)}$ is necessary to be calculated. This reminds us of searching for a more efficient way to obtain the optimal spin squeezing states based on these dynamics controlling procedures. This is also the way discussed in paper \cite{Shen}.

Figures \ref{Fig.tro1.ntc&non_ntc} and \ref{Fig.tro2.ntc&non_ntc} show the numerical time evolution of the corresponding two schemes depicted in Fig. \ref{Fig.tro1&2.Pulse}(a) and Fig. \ref{Fig.tro1&2.Pulse}(b). One can see that in both cases, the numerical results exhibit oscillation behaviors away from the ideal dynamics when $\chi t$ is large. For the scheme based on the $1$st-order TS expansion, the spin squeezing parameter $\xi^2$ at time instant $M t_c^{(A)}$ is always on the top of the evolution curve, even for large $N_c$. Therefore, to achieve the ideal spin squeezing at time $t_{opt}$, $\delta t$ should be sufficiently small, which indicates a relatively large $N_c$. However, for scheme A based on the $2$nd-order TS expansion, the corresponding values $\xi^2$ at time instant $M t_c^{(A)}$ moves to the bottom of the evolution curve as $N_c$ increases, as shown in Fig. \ref{Fig.tro2.ntc&non_ntc}. So with much smaller $N_c$, we can obtain a good approximation of the optimal spin squeezing $\xi^2$ by controlling the total evolution time. Figure 5 shows the tracks of $\xi^2$ at time instant $Mt^{(A)}_c$ for different schemes. As a result, we conclude while it requires as many as $N_c=1000$ periods to get a good result using the proposal in \cite{Liu}, a much smaller $N_c$ $(N_c=50)$ is sufficient when scheme A is employed without introducing new controlling pulses.

\begin{figure}
 \includegraphics[width=1.0 \columnwidth]{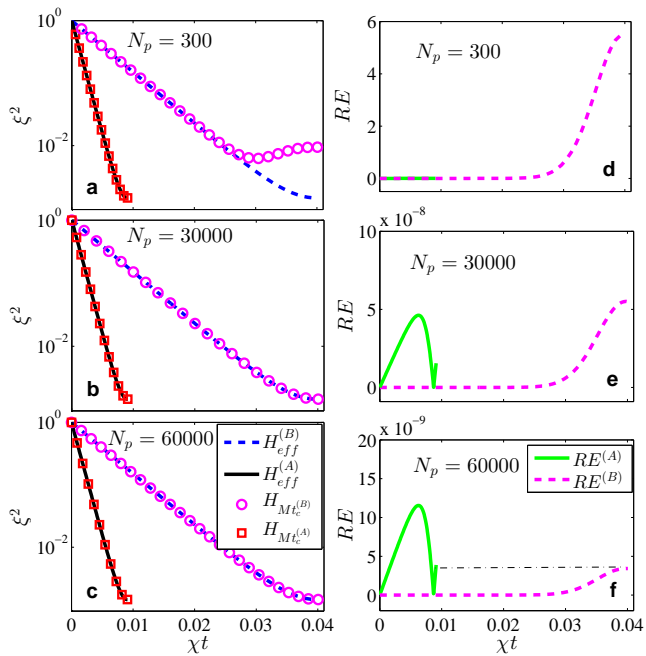}
  \caption{(Color online) The spin squeezing (a, b, c) and associated relative errors (d, e, f). In (a), (b), and (c) the red squares, the solid black line, the magenta circles, and the dashed blue line correspond to the evolution of scheme A, $H_{eff}^{(A)}$, and scheme B, $H_{eff}^{(B)}$, respectively. The solid green line and the dashed magenta line in (d), (e), and (f) are the relative errors of scheme A and B getting from $|\xi_{Mt_c^{(A)}}^{2(A)}-\xi_{eff}^{2(A)}|/\xi_{eff}^{2(A)}$ and $|\xi_{Mt_c^{(B)}}^{2(B)}-\xi_{eff}^{2(B)}|/\xi_{eff}^{2(B)}$.}
\label{Fig.tro1&2.error}
\end{figure}

To investigate the efficiency of the scheme based on higher-order TS expansions, in Fig. \ref{Fig.tro3.ntc} we plot the squeezing parameter following scheme B for $N=1250$ with pulse number $N_p=100$. One can see that a large deviation from the ideal evolution derived by $H_{eff}^{(B)}$ appears as the duration time grows. In principle, when the duration time of the single period of TS expansion is fixed, for the same evolution time, the higher-order TS expansion will lead to the higher precision compared with the lower-order one. However, in this problem our scheme B takes a longer evolution time to achieve the optimal squeezing compared with scheme A. With the effective Hamiltonian $\chi(J_x^2-J_y^2)/(12s-3)$, scheme B needs an evolution time $(12s-3)t_{opt}$ to realize the TAT optimal squeezing,
while scheme A only takes time $3t_{opt}$. So for the fixed pulse number $N_p=100$ in Fig. \ref{Fig.tro1&2.ntc} and \ref{Fig.tro3.ntc}, the duration times of one single period for the two schemes satisfy $(12s-3)t_{opt}/(N_p/N_p^{(B)}) \approx 0.79 t_{opt} \gg 3 t_{opt}/(N_p/N_p^{(A)})=0.06 t_{opt}$.
This is the reason scheme A has a better result.
With the increase of the pulse number $N_p$, we can make the duration times $(12s-3)t_{opt}/(N_p/N_{p}^{(B)})$ and $3 t_{opt}/(N_p/N_{p}^{(A)})$ smaller and smaller, and finally scheme B will have a better result than scheme A because the error in the $3$-rd order expansion decreases faster than that of the $2$nd-order expansion.
Figure \ref{Fig.tro1&2.error} shows the error analysis for both schemes with a different total number of pulses $N_p$.
One can see that scheme A always has a relatively lower error rate and shorter evolution time until $N_p$ reaches $60\,000$. Such a large $N_p$ requires too many resources.
We conclude that scheme B may have higher precision when $N_p$ is large enough, but it requires too many controlling pulses and is experimentally impractical. A simplified scheme A based on $2$nd-order TS expansion is enough for our purposes.

After the paper of Ref.\cite{Liu}, Shen \textsl{et al}. also presented an idea \cite{Shen} to enhance the performance of OAT to get spin squeezed states close to the Heisenberg limit. Compared with their proposal, our result takes a shorter evolution time. Taking $N=2000$ as an example, the time cost of our scheme A(B) is around $0.006/\chi (0.027/\chi)$, shorter than their $0.1/\chi$. Besides, our scheme is also spin number independent.

\section{Conclusion}
In conclusion, we have developed a scheme using a series of rf pulses to transform an OAT to a TAT Hamiltonian. In contrast to the proposal in Ref.\cite{Liu}, our scheme A reduces the pulse number from $N_c=1000$ to 50 for $N=1250$ atoms, which is very experimentally friendly. With the help of the terms $\Omega(t)J_y$ and $G(t)J_x$, pulse sequences designed according to higher order Trotter-Suzuki expansion can be realized. We find that while scheme B can reach optimal spin squeezing with high precision during the whole evolution, it needs too many controlling pulses and is experimentally impractical. We note that our scheme is spin-number independent, and it can be generalized in other systems where only an OAT Hamiltonian \cite{Molmer,Leroux,Ockeloen} is realized. Moreover, compared to the known work \cite{Shen}, our schemes also have a relatively shorter evolution time. Therefore they should be realizable with current techniques, such as those reported in Refs. \cite{Gross} and \cite{Riedel}.

\section{Acknowledgement}
This work was funded by the National Natural Science Foundation of
China (Grant No. 11174270), National Basic
Research Program of China (Grants No. 2011CB921204 and No. 2011CBA00200), the Fund of CAS, and the Research
Fund for the Doctoral Program of Higher Education of China (Grant
No. 20103402110024). Z. W. Z. gratefully acknowledges the
support of the K. C. Wong Education Foundation, Hong Kong.


\begin{references}
\bibitem{Kitagawa} M. Kitagawa and M. Ueda, Phys. Rev. A \textbf{47}, 5138 (1993).
\bibitem{Heinzen} D. J. Wineland, J. J. Bollinger, W. M. Itano, F. L. Moore, and D. J. Heinzen, Phys. Rev. A \textbf{46}, R6797 (1992).
\bibitem{Wineland} D. J. Wineland, J. J. Bollinger, W. M. Itano, and D. J. Heinzen, Phys. Rev. A \textbf{50}, 67 (1994).
\bibitem{Ma} J. Ma, X. G. Wang, C. P. Sun, and F. Nori, Phys. Rep. \textbf{509}, 89 (2011).
\bibitem{Zoller} A. S{\o}rensen, L. M. Duan, J. I. Cirac, and P. Zoller, Nature \textbf{409}, 63 (2001).
\bibitem{Bigelow} N. Bigelow, Nature \textbf{409}, 27 (2001).
\bibitem{Sorensen} A. S. S{\o}rensen and K. M{\o}lmer, Phys. Rev. Lett. \textbf{86}, 4431 (2001).
\bibitem{Korbicz} J. K. Korbicz, J. I. Cirac, and M. Lewenstein, Phys. Rev. Lett. \textbf{95}, 120502 (2005).
\bibitem{Amico} L. Amico, R. Fazio, A. Osterloh, and V. Vedral, Rev. Mod. Phys. \textbf{80}, 517 (2008).
\bibitem{Pezze} L. Pezz\'{e} and A. Smerzi, Phys. Rev. Lett. \textbf{102}, 100401 (2009).
\bibitem{Horodecki} R. Horodecki, P. Horodecki, M. Horodecki, and K. Horodecki, Rev. Mod. Phys. \textbf{81}, 865 (2009).
\bibitem{Guhne} O. G\"{u}hne and G. T\'{o}th, Phys. Rep. \textbf{474}, 1 (2009).
\bibitem{Lee} T. E. Lee and C. K. Chan, Phys. Rev. A \textbf{88}, 063811 (2013).
\bibitem{Bollinger} J. J. Bollinger, W. M. Itano, D. J. Wineland, and D. J. Heinzen, Phys. Rev. A \textbf{54}, R4649 (1996).
\bibitem{Leibfried} D. Leibfried, M. D. Barrett, T. Schaetz, J. Britton, J. Chiaverini, W. M. Itano, J. D. Jost, C. Langer, and D. J. Wineland, Science \textbf{304}, 1476 (2004).
\bibitem{Polzik} E. S. Polzik, Nature \textbf{453}, 45 (2008).
\bibitem{Cronin} A. D. Cronin, J. Schmiedmayer, and D. E. Pritchard,   Rev. Mod. Phys. \textbf{81}, 1051 (2009).
\bibitem{Appel} J. Appel, P. J. Windpassinger, D. Oblak, U. B. Hoff, N. Kjaergaard, and E. S. Polzik, Proc. Natl. Acad. Sci. \textbf{106}, 10960 (2009).
\bibitem{Gross} C. Gross, T. Zibold, E. Nicklas, J. Est\`{e}ve, and M. K. Oberthaler, Nature \textbf{464}, 1165 (2010).
\bibitem{Riedel} M. F. Riedel, P. B\"{o}hi, Y. Li, T. W. H\"{a}nsch, A. Sinatra, and P. Treutlein, Nature \textbf{464}, 1170 (2010).
\bibitem{Lipkin} H. J. Lipkin, N. Meshkov, and A. J. Glick, Nucl. Phys. \textbf{62}, 188 (1965).

\bibitem{Vidal} J. Vidal, G. Palacios, and R. Mosseri, Phys. Rev. A \textbf{69}, 022107 (2004); J. Vidal, G. Palacios, and C. Aslangul, ibid. \textbf{70}, 062304 (2004)

\bibitem{Haine} S. A. Haine and M. T. Johnsson, Phys. Rev. A \textbf{80}, 023611 (2009).
\bibitem{Liu} Y. C. Liu, Z. F. Xu, G. R. Jin, and L. You, Phys. Rev. Lett. \textbf{107}, 013601 (2011).
\bibitem{Shen} C. Shen and L. M. Duan, Phys. Rev. A \textbf{87}, 051801 (2013).
\bibitem{Yin} X. L. Yin, J. Ma, X. G. Wang, and F. Nori, Phys. Rev. A  \textbf{86}, 012308 (2012).
\bibitem{Louchet} A. Louchet-Chauvet, J. Appel, J. J. Renema, D. Oblak, N. Kjaergaard, and E. S. Polzik, New J. Phys. \textbf{12}, 065032 (2010).
\bibitem{Julia} B. Juli\'{a}-D\'{\i}az, T. Zibold, M. K. Oberthaler, M. Mel\'{e}-Messeguer, J. Martorell, and A. Polls, Phys. Rev. A \textbf{86}, 023615 (2012).
\bibitem{Chaudhry} A. Z. Chaudhry and J. B. Gong, Phys. Rev. A \textbf{86}, 012311 (2012).
\bibitem{Suzuki} M. Suzuki, J. Math. Phys. \textbf{32}, 400 (1991).
\bibitem{Molmer} K. M{\o}lmer and A. S{\o}rensen, Phys. Rev. Lett. \textbf{82}, 1835 (1999).
\bibitem{Leroux} I. D. Leroux, M. H. Schleier-Smith, and V. Vuleti\'{c}, Phys. Rev. Lett. \textbf{104}, 073602 (2010).
\bibitem{Ockeloen} C. F. Ockeloen, R. Schmied, M. F. Riedel, and P. Treutlein, Phys. Rev. Lett. \textbf{111}, 143001 (2013).

\end{references}
\end{document}